\documentclass[12pt]{iopart}

\usepackage{hyperref}
 
\usepackage[dvips]{graphicx}
 
\def\v#1{{\bf#1}}
\def\be{\begin{equation}}
\def\ee{\end{equation}}
\def\bea{\begin{eqnarray}}
\def\eea{\end{eqnarray}}
\def\ahalf{{\textstyle{1\over2}}}

\newcommand{\bfbeta}{\mbox{\boldmath$\beta$\unboldmath}}

\newcommand{\bfsigma}{\mbox{\boldmath$\sigma$\unboldmath}}

\def\ie{{\it i.e.\,}}
\def\etal{{\it et al.   }}

\def\pcaliph{\mbox{$\cal P\,$}}
\def\<{\langle}
\def\>{\rangle}
 
\begin{document}

\title{Klein-Gordon and Dirac gyroscopes}
 
 
\author{E Sadurn\'i $^1$ }
 
\address{$^1$Instituto de Ciencias F\'isicas,
Universidad Nacional Aut\'onoma de M\'exico,
Cuernavaca, Morelos, M\'exico.}
 
 
 
 
\ead{sadurni@fis.unam.mx}

 
\begin{abstract}
The formulation of a rigid body in relativistic quantum mechanics
is studied. Departing from an alternate approach at the relativistic classical level, the corresponding Klein-Gordon and Dirac operators for the rigid body are obtained in covariant form. The resulting wave equations are shown to be consistent, by construction, with earlier definitions of a relativistic rigid body by Aldinger \etal (1983). Wave functions and spectra for both cases are obtained explicitly, including the  Dirac gyroscope with asymmetries.
 
\end{abstract}
 
\submitto{ \JPA }
 
 
\maketitle

\section{Introduction}

The study of relativistic rigid bodies in quantum mechanics has been explored extensively in previous works, particularly in \cite{aldinger}. Relativistic quantum rigid bodies (RQRB) are defined through a set of three fundamental properties: 1) elementary limit, 2) consistent classical limit, 3) consistent non-relativistic limit. The second requirement, however, leaves some freedom to what a relativistic rigid body should be at the classical level. See, for instance \cite{hanson} and \cite{starus} for some definitions of rotating objects. The interest of the present work is to formulate relativistic wave equations describing rigid bodies of spin $0,\ahalf$ taking advantage of this point (here, spin should not be confused with total angular momentum, sometimes referred to as total spin). Since the focus is centered on rigid bodies at subatomic scales, general relativity effects will be neglected. 

A construction of simple wave equations describing RQRB is persuited. To achieve this, we find it convenient to formulate classical relativistic gyroscopes as multiparticle systems. Moreover, a manifestly Lorentz invariant construction is possible when such particles interact with each other through Lorentz covariant gauge potentials. Any effective realization of a rigid body in nature must be through such interactions, though special conditions are needed for this to happen. Our alternate approach to a classical rigid body will be established by means of an analogy between non-relativistic and relativistic expressions for the energy and the squared energy respectively. This will ensure an appropriate non-relativistic limit in our treatment. It must be mentioned that constructions of relativistic wave equations for many particles have been given before, for example, in \cite{add4}, \cite{add5}. In connection with relativistic classical mechanics, the problems of orbit reconstruction, one-time description, separability of rotational kinematics and appropriate definitions of center of mass in many body relativistic systems have been treated extensively in \cite{add6}, \cite{add7}. The study of these complex problems entails special definitions and generalizations of non-relativistic concepts (e.g. dynamical body frames, canonical spin bases, canonical internal center of mass, etc.) which, however, are not used in the present work. The $N$-body approach presented here uses gauge interactions as an instrument to ensure Lorentz covariance. Moreover, our covariant definition of center of mass (indicated later in the text) is motivated by its properties when replaced in quadratic forms of particles' coordinates and momenta. Although important results have been achieved in the references cited above, we may instead use the strategy of identifying non-relativistic expressions related to rigid bodies through manipulations within the relativistic framework.

As a motivation for the present work, it should be mentioned that (non-relativistic) rotational behavior
is highly important in the study of molecular \cite{pina} and nuclear spectra \cite{wadsworth}. For subnuclear structures arising in the realm of high energy physics, a relativistic treatment should be useful. Hadronic spectroscopy
in the absence of vibrational modes has been explored before in this context \cite{macgregor} and results of the present paper may be applied to this ground as well. Alternatively, there has been interest in the study of kicked rotators and their implications to quantum chaos \cite{seligman}, reaching recently the relativistic realm in \cite{matrasulov}.

The paper is structured as follows. In section 2 the classical relativistic treatment is established through a system of interacting particles. It includes a suitable definition for a relativistic inertia tensor. In section 3 the corresponding Klein-Gordon operator is written. Wavefunctions and energies are obtained. Section 4 shows the procedure to obtain a Dirac equation for RQRB. Here, inertia tensor with a real diagonal inverse square root is considered. Wave functions and energies for the symmetric case are written explicitly. Section 5 contains yet another way of writing the Dirac operator with a non-abelian inverse square root for the inertia tensor, though moments of inertia remain the same as their classical counterparts. Through this treatment, analytical expressions for wave functions and energies are given in simple form for the case of asymmetric spin-orbit coupling. Section 6 contains a brief conclusion.

\section{An alternative approach to a relativistic rigid body in classical mechanics}

In the formulation of the free rigid body problem, one could follow a path well depicted in classical textbooks \cite{landau}, by considering the system to comprise many particles obeying the holonomic constraints of fixed distances between each pair of them. The velocities of the constituents in any frame of reference can be written in terms of a displacement velocity and an angular velocity which is the same for all points. Such a treatment, when employed in the non-relativistic realm, leads to a kinetic energy (and hamiltonian for the free case) in the center of mass of the form

\be
H = \ahalf \sum_{i=1}^{3} I_i^{-1} L_i^2 = \ahalf \sum_{i=1}^{3} I_i \Omega_i^2
\label{1.1}
\ee
where $\Omega_i$ are the components of the angular velocity, $L_i$ are the components of the total angular momentum in the body-fixed frame and $I_i$ are the moments of inertia. In the special theory of relativity, however, the form of the kinetic energy allows neither a quadratic term in the velocities (Lorentz factors are present through the relativistic mass) nor a simple decomposition in terms of displacement and angular velocity with respect to some origin. Even in the case when such an origin is chosen as the center of mass and at rest, a definition of inertia tensor is possible only when the defined object depends on the velocity of each particle by means of the corresponding Lorentz factors. The general description becomes rather involved with no parallel to the simpler form (\ref{1.1}) unless the non-relativistic limit is taken. It should be noticed that despite of this, energy and angular momentum are constants of the motion when viewed in the frozen center of mass - a one-time description is possible in this frame of reference even at the quantum domain (see \cite{nikitin}) and rotational invariance also holds.

With this in mind, we start our approach with a system of $N$ particles (though this number is irrelevant) interacting through gauge potentials coming from a Lorentz covariant theory. Ultimately, it is through these that a rigid body is effectively realized in nature. Let $r^{(l)}_{\mu}$ be the four-vector representing the coordinates of the $l$-th particle, $p^{(l)}_{\mu}$ its canonically conjugate four-momentum, $A^{(l)}_{\mu}$ the gauge potential felt by the particle due to other components of the system, $m_l$ its rest mass and $\pi^{(l)}_{\mu}$ the corresponding mechanical four-momentum such that

\bea
\pi^{(l)}_{\mu} &=& \left(\frac{m_l c}{\sqrt{1-(v^{(l)}/c)^2}}\right) \frac{d r^{(l)}_{\mu}}{d r^{(l) 0}} \nonumber \\
                &=&  p^{(l)}_{\mu}-A^{(l)}_{\mu}
\label{0.1}
\eea
with $(v^{(l)})^2 = c^2(d \v r^{(l)}/ d r^{(l)}_{0})^2$. The energy of the particle can be written as

\bea
E=cp^{(l) 0}=\sqrt{c^2(\v p^{(l)}- \v A^{(l)})^2+m_l^2c^4}+cA^{(l) 0}
\label{0.2}
\eea
which is equivalent to the Lorentz invariant expression

\be
\pi^{(l)}_{\mu} \pi^{(l) \mu} = - (m_l c)^2
\label{1.2}
\ee
where the sum convention over repeated greek indices is adopted. Considering a transformation to the center of mass and relative coordinates, \ie the Jacobi transformation \cite{marcosbook}, we write the resulting set of coordinates and their conjugate momenta denoted by a dot above in the form

\bea 
\fl
\dot r^{(N)}_{\mu} \equiv R_{\mu} = \sum_{i=1}^{N} \sqrt{\frac{ m_i}{N M}} r^{(i)}_{\mu}, &\fl{\rm center \, of \, mass \ }& \nonumber \\ \fl \dot r^{(l)}_{\mu} = \frac{1}{\sqrt{l(l+1)}} \sum^{l}_{i=1}\left( \sqrt{\frac{m_i}{M}} r^{(i)}_{\mu} -\sqrt{\frac{m_{l+1}}{M}} r^{(l+1)}_{\mu} \right), &\fl{\rm relative \, coordinates \ }& \nonumber \\
\fl \dot p^{(N)}_{\mu} \equiv P_{\mu} = \sum_{i=1}^{N} \sqrt{\frac{M}{N m_i}} p^{(i)}_{\mu}, &\fl{\rm center \, of \, mass \, momentum \ }& \nonumber \\ \fl \dot p^{(l)}_{\mu} = \frac{1}{\sqrt{l(l+1)}} \sum^{l}_{i=1}\left( \sqrt{\frac{M}{m_i}} p^{(i)}_{\mu} -\sqrt{\frac{M}{m_{l+1}}} p^{(l+1)}_{\mu} \right),  &\fl{\rm relative \, momenta \ }& \nonumber \\  \fl
\dot \pi^{(l)}_{\mu} \equiv \dot p^{(l)}_{\mu}- \dot A^{(l)}_{\mu}, \qquad \dot A^{(l)}_{\mu} = \frac{1}{\sqrt{l(l+1)}} \sum^{l}_{i=1}\left( \sqrt{\frac{M}{m_i}} A^{(i)}_{\mu} -\sqrt{\frac{M}{m_{l+1}}} A^{(l+1)}_{\mu} \right) \nonumber \\ \fl l=1,...,N-1
\label{1.3}
\eea
where $M$ is the total rest mass. Notice that index $l$ above distinguishes Jacobi coordinates and is not related to particles. Since we are dealing with a free RQRB, there is no need to define a mechanical momentum for the center of mass: despite the transformed vector potential for $l=N$ could be non-vanishing, it is necessarily independent of $R_{\mu}$
and can be gauged away. Multiplying relation (\ref{1.2}) by $M/m_l$ and summing over $l$ we obtain

\bea
\sum_{l=1}^{N} \frac{M}{m_l} \pi^{(l)}_{\mu} \pi^{(l) \mu} = -(Mc)^2
\label{1.4}
\eea
or

\bea
P_{\mu}P^{\mu} + \sum_{l=1}^{N-1} \dot \pi^{(l)}_{\mu} \dot \pi^{(l) \mu} = -(Mc)^2
\label{1.5}
\eea
which follows from the fact that Jacobi coordinates come from a rotation in particle space. Furthermore, we can separate space-like and time-like terms in the form $\pi_S^2 \equiv \sum_{l=1}^{N-1} \dot \pi^{(l)}_{j} \dot \pi^{(l) j}$, $\pi_T^2 \equiv \sum_{l=1}^{N-1} \dot \pi^{(l)}_{0} \dot \pi^{(l) 0}$ such that

\bea
P_{\mu}P^{\mu} + \pi_S^2 + \pi_T^2 = -(Mc)^2
\label{1.6}
\eea
and the energy on the center of mass ($P^{i}=0$) can be cast as a function of Jacobi mechanical momenta, \ie

\bea
cP^{0}  = \sqrt{c^2\pi_S^2 + c^2\pi_T^2 + M^2c^4}.
\label{1.7}
\eea
Alternatively, one could solve for $p^{(l)}_{0}$ in (\ref{1.2}), multiply by $\sqrt{M/m_l}$ and sum over $l$ to obtain the same energy in terms of particle coordinates and mechanical momenta

\bea
c P^0 = \sum_{l=1}^{N} \sqrt{M/m_l} \left( \sqrt{c^2 \pi^{(l)}_{j} \pi^{(l) j} + m_l^2c^4} + cA^{(l)0}\right).
\label{1.8}
\eea
Now we turn to the crucial point. Let $V=\sum_{l=1}^{N} A^{(l) 0}$. Had we formulated our problem in the non-relativistic domain, a hamiltonian of the form

\bea
H = \sum_{l=1}^{N} \frac{\pi^{(l)}_{j} \pi^{(l) j}}{2m_l} + V = \frac{\pi_S^2}{2M} + V
\label{1.9}
\eea
would have appeared. When vector potentials are responsible for an effective realization of a rigid body, hamiltonians (\ref{1.1}) and (\ref{1.9}) can be identified. This, together with the second equality in (\ref{1.9}), allows to write $\pi_S^2$ in terms of $L_i, I_i$ and $V$. When the resulting $\pi_S^2$ is replaced in (\ref{1.6}) we get

\bea
-(P^{0})^2 + M \sum_{i=1}^{3} I_i^{-1} L_i^2 + \pi_T^2 - 2MV = -(Mc)^2,
\label{1.10}
\eea
and through this treatment we have made appear the familiar quadratic form in the angular momentum. The time-like part $\pi_T^2$ is thus far unknown. In order to deal with this term, we may use the functional form of (\ref{1.10}) together with the conservation of energy and angular momentum at the center of mass coming from the invariance properties of (\ref{1.8}). Conservation laws applied to (\ref{1.10}) imply
$\pi_T^2 - 2MV = const. \equiv \pcaliph^2 $, which may depend on angular momentum and the body's rest energy - there are no more independent conserved quantities available in principle. The non-relativistic limit $Mc^2 \rightarrow \infty $ and the full relativistic limit $Mc^2 \rightarrow 0 $ are helpful at this point. The first limit applied in (\ref{1.10}) leads to

\be
cP^0 \approx Mc^2 + \ahalf \sum_{i=1}^{3} I_i^{-1} L_i^2 + \frac{\pcaliph^2}{2M}.
\label{1.11}
\ee
However, at order $1/M$ we expect the energy to be given exclusively by the first two terms in the r.h.s. of (\ref{1.11}). Thus $\pcaliph^2 \sim 1/M$ or higher inverse powers. Using this result, the second limit would lead to an infinite energy in the absence of rest mass, which makes no sense for a system of interacting particles. Therefore we must choose $\pcaliph^2=0$ to meet both requirements and the energy is then given by

\be
E= \sqrt{Mc^2\sum_{i=1}^{3} I_i^{-1} L_i^2 + (Mc^2)^2}.
\label{1.12}
\ee
It should be emphasized that $I_i$ are moments of inertia at the center of mass which coincide with
the usual non-relativistic definitions \cite{landau}. 

\subsection{Classical equation in Lorentz invariant form}

Since we want (\ref{1.12}) to be manifestly Lorentz invariant, we proceed to give a covariant definition for the inertia tensor. First of all, we need a $4 \times 4$ symmetric tensor which is diagonal in some basis. Moreover, (\ref{1.12}) indicates it must reduce to 

\bea
I_{ij} = \sum_{l=1}^{N} m_l \left( r_i^{(l)} r_j^{(l)} - \delta_{ij} (\v r^{(l)})^2 \right)
\label{1.13}
\eea
at the center of mass, while in other frames of reference differing by a constant velocity, $I_{ij}$ should account for a distortion of the body in the direction of a boost. Thus we define 

\bea
u_{\mu} = \frac{P_{\mu}}{\sqrt{-P^{\nu}P_{\nu}}} \nonumber \\
r^{(l)}_{\perp \mu} = r^{(l)}_{\mu} - u_{\mu} ( u^{\nu} r^{(l)}_{\nu} ) \nonumber \\
I_{\mu \nu} = \sum_{l=1}^{N} m_l \left( r_{\perp \mu}^{(l)} r_{\perp \nu}^{(l)} - \eta_{\mu \nu}  r^{(l)}_{\perp \lambda} r^{(l) \lambda}_{\perp} \right)
\label{1.14}
\eea
where $\eta$ denotes the metric tensor. For later use we define also the covariant object $\bar I_{\mu \nu}$ such that $\bar I_{\mu \rho} \bar I^{\rho}_{\, \, \nu} = (I^{-1})_{\mu \nu} $.

If (\ref{1.10}) is written as a Lorentz scalar in the appealing form

\bea
P_{\mu}P^{\mu} + B_{\mu}B^{\mu} + (Mc)^2 = (P+B)_{\mu}(P+B)^{\mu} + (Mc)^2 = 0
\label{1.15}
\eea
then it will be necessary to find $B_{\mu}$ such that, at the center of mass, the term  $B_{\mu}B^{\mu}$ reduces to the kinetic energy of the rigid body and $B_{\mu}P^{\mu}$ vanishes for any frame of reference. For this we find it useful to define an analog of the Pauli-Lubanski vector \cite{ryder} for a system of particles in terms of the angular momentum antisymmetric tensor, \ie

\bea
M_{\mu \nu} = \sum_{l=1}^{N} r^{(l)}_{\mu} p^{(l)}_{\nu} - \mu \leftrightarrow \nu =\sum_{l=1}^{N} \dot r^{(l)}_{\mu} \dot p^{(l)}_{\nu} - \mu \leftrightarrow \nu \nonumber \\
W_{\mu} = -\frac{1}{2} \epsilon_{\mu \nu \sigma \rho} P^{\nu} M^{\sigma \rho}
\label{1.16}
\eea 
where $\epsilon$ is the totally antisymmetric symbol, $\epsilon_{0123}=1$. The sum in the angular momentum may not include the $N$th term, since it vanishes when replaced in the expression for $W$. With these definitions, it is straightforward to show that 

\be
B_{\mu} = \sqrt{\frac{M}{-P^{\nu}P_{\nu}}} \bar I_{\mu}^{\, \, \rho} W_{\rho} =  \frac{ \sqrt{M}}{2} \epsilon_{\nu \sigma \lambda \rho} u^{\nu} M^{\sigma \lambda} \bar I_{\mu}^{ \, \, \rho}
\label{1.17}
\ee
does the desired job, as long as $L_i = \ahalf \epsilon_{ijk}L_{jk}$ with $L_{jk}$ the body fixed angular momentum \ie when $I_{jk}$ is diagonal. Thus (\ref{1.15}) and (\ref{1.17}) are the sought invariant expressions for the classical RRB.

\section{The Klein-Gordon gyroscope}

By following the canonical quantization prescription, namely $p^{(l)}_{\mu} \rightarrow -i\hbar \partial^{(l)}_{\mu} $, it results very simple to promote (\ref{1.15}) and (\ref{1.17}) to operator form. Nevertheless, moments of inertia remain as parameters, not operators. We have thus the equation

\be
\left[ \left(P + B \right)_{\mu} \left(P + B \right)^{\mu} + (Mc)^2 \right] \phi = 0
\label{2.1}
\ee
where $\phi=\phi(r^{(l)}_{\mu})$. In terms of the D'Alembertian $\triangle$ with respect to $R_{\mu}=(ct,\v R)$
we have

\be
\left[ -\hbar^2 \triangle + B_{\mu}B^{\mu} + (Mc)^2 \right] \phi = 0.
\label{2.2}
\ee
To compute the energies and wave functions at the center of mass, we replace $u_{\mu}=(-1,0,0,0)$ to get

\be
\left[Mc^2\sum_{i=1}^{3} I_i^{-1} L_i^2 + M^2c^4 + \hbar^2 \frac{\partial^2}{\partial t^2} \right]\phi=0
\label{2.3}
\ee
or its stationary version

\be
\left[Mc^2\sum_{i=1}^{3} I_i^{-1} L_i^2 \right]\phi=\left(E^2-M^2c^4\right)\phi.
\label{2.4}
\ee
The eigenvalues of the operator in (\ref{2.4}) determine the energy and can be computed for the general (asymmetric) case by means of algebraic methods \cite{draayer} or direct diagonalization using appropriate states \cite{wang}. In fact, it is suitable to use kets $|lm\>$ such that $m$ indicates the angular momentum projection in the body frame (in the laboratory frame such number is always conserved, but it is not customary to include it in the notation), while $l$ is the total angular momentum number in the body frame. It is a conserved quantity, since $L^2$ commutes with the operator in (\ref{2.4}). Restricting ourselves to the symmetric case $I_1=I_2$, eigenvalues are

\be
E_{lm}= \pm \sqrt{Mc^2 \hbar^2 \left[ I^{-1}_1 l(l+1) + \left( I^{-1}_3-I^{-1}_1 \right)m^2 \right] + M^2c^4 }
\label{2.5}
\ee
and wave functions are those of the non-relativistic problem. Here we could equally follow the approach in \cite{pina}, where the asymmetry is treated by three restricted parameters given in terms of the moments of inertia to obtain analytical solutions using the same states $|lm\>$.

Finally, it must be mentioned that requirements 1), 2), 3) in \cite{aldinger} are fulfilled by construction. Specifically, 1) is reached by setting $N=1$ in (\ref{2.1}), for which $B_{\mu}=0$ from its definition and $P_{\mu}$ becomes the single particle momentum. Requirement 2) is obviously met, since (\ref{1.15}) and (\ref{1.17}) were our points of departure for quantizing the system. Equation (\ref{1.11}) with $\pcaliph^2=0$ shows 3).

\section{The Dirac gyroscope}

The need for a relativistic equation which is linear in the generator of time translations may lead us to a generalization of the Dirac equation motivated by (\ref{2.1}). In such case, an intrinsic spin would appear as a consequence of the corresponding Clifford algebra which may be related to the spin of individual particles comprised by the RQRB. For requirement 1) to be fulfilled, the spin of the particles should not contribute to the energy, as it is the case for a single particle obeying the Dirac equation. Thus, a Dirac equation which is linear in $P_{\mu}$ and $B_{\mu}$ can be proposed in the form

\bea
\left[ \gamma_{\mu}\left( P^{\mu}+B^{\mu} \right) + Mc \right] \psi = 0
\label{3.1}
\eea
where $\gamma_{\mu}$ are Dirac matrices \cite{dirac} and $B_{\mu}$ is taken as in (\ref{1.17}), \ie without spin contribution. Equation (\ref{3.1}) is motivated by what would be the {\it square root\ } of (\ref{2.1}). We shall refer to it as the Dirac gyroscope. At the center of mass, (\ref{3.1}) possesses a hamiltonian form given by

\bea
H\psi= \left[ \sqrt{M}c \alpha \cdot \left(\bar {\rm I\ } \v L \right) + \beta Mc^2 \right]\psi=i\hbar\frac{\partial \psi}{\partial t}
\label{3.2}
\eea
with $\alpha_{i}=\gamma^0 \gamma_{i} = \beta \gamma_{i}$ and $\bar {\rm I\ } = \rm{diag} \lbrace I_1^{-1/2},I_2^{-1/2},I_3^{-1/2}\rbrace $ in the frame of principal axes. Squaring (\ref{3.2}) and using the relations

\bea
S_i = \frac{\hbar}{2} \sigma_i \otimes \v 1_{2}, \qquad \alpha_i \alpha_j = \v 1_4 \delta_{ij} + \frac{2i}{\hbar} \epsilon_{ijk}S_k \nonumber \\
L_i L_j = \frac{1}{2} \lbrace L_i , L_j \rbrace + \frac{i \hbar}{2} \epsilon_{ijk}L_k
\label{3.4}
\eea
yields

\bea
\left[ Mc^2\sum_{i=1}^{3} I_i^{-1} L_i^2 - Mc^2\sum_{i=1}^{3} I_i^{-1} L_i S_i + M^2c^4 \right] \psi = -\hbar^2 \frac{\partial^2 \psi}{\partial t^2}
\label{3.5}
\eea
which resembles closely (\ref{2.3}), but with a spin-orbit coupling term containing asymmetries through $\bar {\rm I\ }$. Here, the situation is quite similar to what ocurrs in the case of the Dirac oscillator \cite{DO} when its square energy is computed. The Dirac hamiltonian (\ref{3.2}) describes a RQRB with an extra term appearing in phenomenological hamiltonians for mass operators in nuclear \cite{bohr} and hadron physics \cite{sadur}. Requirement 3) is fulfilled again as long as we allow our non-relativistic gyroscope to contain spin-orbit coupling. Requirements 1) and 2) are met by the same reasons exposed for the Klein-Gordon gyroscope. In the classical limit, the spin-orbit term disappears in (\ref{3.5}).

Now we turn to the solutions of (\ref{3.2}) in its stationary version, which can be discussed through (\ref{3.5}) or by using a more instructive approach. We follow the latter and present the spherical ($I_i=I$) and symmetrical ($I_1=I_2$) cases separately, although one reduces to the other.

\subsection{Spherical case}

First of all, it should be noticed that the asymmetric case is such that $\v J \equiv \v L + \v S$ is not a conserved quantity, since it does not stand for the total angular momentum. Instead, it is the sum of the laboratory spin and the body-fixed orbital angular momentum. Since our problem is rotationally invariant in general, states are determined by the total angular momentum in the laboratory frame and its projection in the quantization axis. However, we omit the state dependence on such quantum numbers as we did for the laboratory angular momentum projection in the Klein-Gordon case (in fact, this applies even for the non-relativistic case \cite{draayer}). States labeled $|lm\>$ will be used, but as it can be noticed, $\v J$ does commute with $H$ in the spherical case and we may use states $|j(l,\ahalf) m_j \>$ to find the spectrum. The spherical hamiltonian reduces to

\bea
\fl H = \frac{2 c }{\hbar} \sqrt{\frac{M}{I}} \beta_1 (\v L \cdot \v S) + \beta_3 Mc^2 = \frac{ c }{\hbar} \sqrt{\frac{M}{I}} \beta_1 \left( J^2-L^2-\frac{3 \hbar^2}{4} \right) + \beta_3 Mc^2
\label{3.6}
\eea
where $\beta_i = \v 1_2 \otimes \sigma_i $, with the properties $\left[ \beta_i , S_j \right]=\left[ \beta_i , L_j \right]=0$, $\lbrace \beta_i , \beta_j \rbrace = \v 1_4 \delta_{ij}$. Because of this, Dirac states for this system can be factorized in the form $| j(l,\ahalf) m_j \> \otimes \chi$, with $\chi$ a Pauli spinor whose up and down components stand for the big and small parts of the wave functions. Therefore, solutions read

\bea
\fl \psi_{jlm_j \pm} =  |j(l,\ahalf) m_j \>  \otimes \chi_{\pm} \nonumber \\
\fl E_{jl, \pm} = \pm \sqrt{ \frac{\hbar^2 Mc^2 }{ I } \left( j(j+1)-l(l+1)-\frac{3}{4} \right)^2 + M^2c^4  }
\label{3.7}
\eea
where $j=l \pm \ahalf$. Spinor $\chi_{\pm}$ obeys the algebraic equation

\bea
\fl \left[c \hbar \sqrt{\frac{M}{I}} \left( j(j+1)-l(l+1)-\frac{3}{4} \right) \sigma_1 +  Mc^2 \sigma_3 \right] \chi_{\pm} = E_{jl, \pm} \chi_{\pm}
\label{3.8}
\eea
which can be solved in terms of the canonical basis $|\pm\>$, \ie

\be
\chi_{\pm} = \sqrt{\frac{E_{jl, \pm}+Mc^2}{2E_{jl, \pm}}} |+\> + \sqrt{\frac{E_{jl, \pm}-Mc^2}{2E_{jl, \pm}}} |-\>.
\label{3.9}
\ee

\subsection{Symmetric case}

Here we define $c_1 \equiv (c/\hbar)\sqrt{M/I_1}, c_3 \equiv (2c/\hbar)\sqrt{M/I_3}$ and take $S_{\pm}, L_{\pm}$ as the standard ladder operators for spin and orbital angular momentum. Hamiltonian (\ref{3.2}) is written as

\bea
H= \beta_1 \left[ c_1 \left( S_{+}L_{-}+S_{-}L_{+}  \right) + c_3 S_3 L_3 \right] + \beta_3 Mc^2 \equiv \beta_1 K + \beta_3 Mc^2
\label{3.10}
\eea
Clearly, $L^2$ and $J_3$ are commuting integrals of the motion. Eigenstates are again separated in the form $\psi = \phi \otimes \chi$, where $\phi$ is labeled by $m_j$ (the $z$ projection of $\v J$) and $l$. Again, spinor $\chi$ contains big and small components of the wave function. Solutions are obtained by replacing these states in the stationary Schroedinger equation with hamiltonian (\ref{3.10}) and solving a $2 \times 2$ secular equation. The two resulting roots will be labeled $i=1,2$. Results for the operator $K$ are

\bea
\fl K \phi^{i}_{l,m_j} = F^{i}_{l m_j} \phi^{i}_{l,m_j}, \qquad i=1,2 \nonumber \\
\fl \phi^{i}_{l,m_j}= \sqrt{\frac{ 2F^{i}_{l m_j} + \hbar^2 c_3 (m_j+\ahalf) }{ 4F^{i}_{l m_j} + \hbar^2 c_3 }} |l,m_j-\ahalf\>|\ahalf\> + \sqrt{\frac{2F^{i}_{l m_j} - \hbar^2 c_3 (m_j-\ahalf) }{ 4F^{i}_{l m_j} + \hbar^2 c_3 }} |l,m_j+\ahalf\>|_{^{-}}\ahalf \> \nonumber \\
\fl F^{i}_{l m_j} = -\frac{\hbar^2}{4} \left[ c_3 + (-)^i \sqrt{c_3^2 + 4c_1^2 l(l+1) + 4(c_1^2-c_3^2)(m_j^2-1/4)}  \right]
\label{3.11}
\eea
while the energy spectrum and eigenstates result in

\bea
\fl H \psi^{i}_{l, m_j,\pm} = E^{i}_{l m_j,\pm} \psi^{i}_{l, m_j,\pm} \nonumber \\
\fl \psi^{i}_{l, m_j,\pm} = \phi^{i}_{l,m_j} \otimes \chi_{\pm} \nonumber \\
\fl E^{i}_{l m_j,\pm} = \pm \sqrt{ \left( F^{i}_{l m_j} \right)^2 + M^2c^4} \nonumber \\
\fl \chi_{\pm} = \sqrt{\frac{E^{i}_{l m_j, \pm}+Mc^2}{2E^{i}_{l m_j, \pm}}} |+\> + \sqrt{\frac{E^{i}_{l m_j, \pm}-Mc^2}{2E^{i}_{l m_j, \pm}}} |-\>
\label{3.12}
\eea
In the $c_1=c_3$ limit (spherical case), we see that the number $(-)^{i}$ is related to the choice $j=l+(-)^{i} \ahalf$. It must be mentioned that the asymmetric case cannot be solved through these techniques alone, but demands a crude matrix diagonalization. However, there remains the question of whether a Dirac gyroscope governed by three independent parameters possesses an analitically solvable spectrum. This fact takes us to the following section

\section{ The symmetric Dirac gyroscope with non-abelian parameters  }

Now we address the problem of formulating a Dirac gyroscope with three independent parameters and whose stationary equation allows analytical solutions. We do not expect the asymmetric problem to be solvable in simple form, but we can propose a symmetric Dirac gyroscope with {\it asymmetric\ } spin-orbit coupling. In defining a Dirac operator like (\ref{3.1}), there is certain freedom in choosing the square root of the inverse inertia tensor. In fact, if a kinetic term of the form $(\bar {\rm I\ } \v L)^{\dagger}( \bar {\rm I\ } \v L) = \v L \cdot ( {\rm I\ }^{-1} \v L)$ is sought in the second order equation (\ref{3.5}), the relation $\bar {\rm I\ }^{\dagger} \bar {\rm I\ }= {\rm I\ }^{-1}$ can be fulfilled in many ways. It must be noticed, however, that the construction of the Dirac operator requires the tensor $\bar {\rm I\ }$ to be independent of $\v S$ and $\v L$. Thus, a dependence on $\beta_i$ remains as the only possibility, since it is the only non-abelian structure left which commutes with $\v L$ and $\v S$. By the relations written immediately after (\ref{3.6}), we see that the $\beta$'s correspond to an independent observable analogous to spin but related to big and small components of bispinors (we have used this fact in the past sections). We can choose the tensor $\bar {\rm I\ }$ up to rotations as

\bea
\bar {\rm I\ } = \beta_1 \bfbeta \cdot \hat v  {\rm I\ }^{-1/2} = \beta_1 \bfbeta \cdot \hat v \left( \begin{array}{ccc} I_1^{-1/2} & 0 & 0 \\ 0 & I_1^{-1/2} & 0 \\ 0 & 0 & I_3^{-1/2} \end{array} \right)
\label{3.13}
\eea
where $\hat v$ is a unit vector and $\bar {\rm I\ }^{\dagger} \bar {\rm I\ }= {\rm I\ }^{-1}$ can be easily verified. Replacing it in hamiltonian (\ref{3.2}) we get the simple expression

\bea
H=  \bfbeta \cdot \hat v \left[ c_1 \left( S_{+}L_{-}+S_{-} L_{+} \right) + c_3 S_3 L_3 \right] + \beta_3 Mc^2 \equiv \bfbeta \cdot \hat v K + \beta_3 Mc^2
\label{3.14}
\eea
where $c_1$, $c_3$ are taken as before. The operator $K$ commutes with $H$ and eigenstates are similar to those obtained in the last section. Results are again those in (\ref{3.11}) for the kinetic energy operator, but energies and states satisfy

\bea
\fl H \psi^{i}_{l, m_j,\pm} = E^{i}_{l m_j,\pm} \psi^{i}_{l, m_j,\pm} \nonumber \\
\fl \psi^{i}_{l, m_j,\pm} = \phi^{i}_{l,m_j} \otimes \chi_{\pm} \nonumber \\
\fl E^{i}_{l m_j,\pm} = \pm \sqrt{ \left( F^{i}_{l m_j} \right)^2 + M^2c^4 + 2 v_3 F^{i}_{l m_j}Mc^2} \nonumber \\
\fl \chi_{\pm} = \frac{1}{E^{i}_{l m_j,\pm}}\left[ \bfsigma \cdot \hat v F^{i}_{l m_j} + \sigma_3 Mc^2 \right] |\pm \> 
\label{3.15}
\eea
We can see that the spectrum is controlled by three independent parameters $c_1, c_3, v_3$. The third parameter plays the role of a translation of kinetic energy and a dilatation of the rest mass. To obtain a Lorentz invariant version of (\ref{3.14}) it suffices to replace $\hat v$ by a Lorentz vector $v_{\mu}$  projected orthogonally with respect to $u_{\mu}$, \ie $v_{\perp \mu} = v_{\mu} - u_{\mu} (u^{\nu} v_{\nu})$.
Thus, if $\beta_{0}=\v 1_4$, the replacement $\bar I_{\mu \nu} \rightarrow  \beta^{\rho} v_{\perp \rho} \bar I_{\mu \nu}$ in (\ref{3.1}) does the required job. 

\section{Conclusions}

A Lorentz invariant description of a classical relativistic body has been given in terms of the non-relativistic hamiltonian (\ref{1.1}). With this, a Klein-Gordon equation for a RQRB has been obtained and solved. The extension to a Dirac equation has been also achieved, including the explicit forms of energies and eigenfunctions at the center of mass. To the author's knowledge, the approach to a RQRB presented here is entirely new, as well as the resulting equations and spectra. This work is expected to help in the study of relativistic kicked rotators of a richier structure than the ones already considered \cite{matrasulov}. Detailed analysis of the spectrum and its application to hadronic spectroscopy is the subject of future work.

\ack
 
Author is grateful to T. H. Seligman and C. Jung for useful discussions. This work was supported by DGAPA-UNAM.

\end{document}